\documentclass[authoryear]{FLO_v1}%

\usepackage{graphicx}
\usepackage{upgreek}
\usepackage{multicol,multirow}
\usepackage{amsmath,amssymb,amsfonts}
\usepackage{mathrsfs}
\usepackage{amsthm}
\usepackage[figuresright]{rotating}
\usepackage{appendix}
\usepackage[authoryear]{natbib}
\usepackage{ifpdf}
\usepackage[T1]{fontenc}
\usepackage{newtxtext}
\usepackage{newtxmath}
\usepackage{textcomp}
\usepackage{xcolor}
\usepackage{csquotes}
\usepackage[colorlinks,allcolors=blue]{hyperref}
\definecolor{jourcolor}{cmyk}{1,0.57,0.01,0.38}
\hypersetup{
    colorlinks,%
    citecolor=jourcolor,%
    filecolor=jourcolor,%
    linkcolor=jourcolor,%
    urlcolor=jourcolor
}

\theoremstyle{definition}

\articletype{RESEARCH ARTICLE}

\Year{2026}

\Vol{1}

\citearticle{Francisco J. G. de Oliveira~\textit{et al}}

\begin{document}

\title[Inter-turbine spacing and flow unsteadiness effects on wake-induced blade dynamics]{Inter-turbine spacing and flow unsteadiness effects on wake-induced blade dynamics}

\author[F. J. G. de Oliveira et~al.]{Francisco J. G. de Oliveira{{\href{https://orcid.org/0000-0002-1244-523X}{\includegraphics{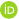}}}}$^{1\ast}$, Adrian T. McGlade$^{1}$, Zahra Sharif Khodaei{{\href{https://orcid.org/0000-0001-5106-2197}{\includegraphics{orcid_logo}}}}$^{1}$ and Oliver R. H. Buxton{{\href{https://orcid.org/0000-0002-8997-2986}{\includegraphics{orcid_logo}}}}}

\address[1]{Department of Aeronautics, Imperial College London, London, UK}

\corres{*}{Corresponding author. E-mail:
\emaillink{f.oliveira22@imperial.ac.uk}}

\keywords{Wind turbines; wakes; wind-farms; fibre-optic measurements}

\date{\textbf{Received:} XX 2020; \textbf{Revised:} XX XX 2020; \textbf{Accepted:} XX XX 2020}

\abstract{Wind turbines operating downstream of others in a farm are routinely exposed to waked inflow, with reduced mean velocity and elevated turbulence driving power deficits and additional structural fatigue. The direct effect of wakes on blade-level structural loading remains under-explored experimentally, owing partly to the sparse spatial coverage of conventional point-based strain sensors. Here, we present a wind-tunnel study of wake-induced blade dynamics using two $1\,\mathrm{m}$-diameter turbine models, in which one blade of a downstream turbine ($WT_2$) is instrumented with distributed Rayleigh-backscattering fibre-optic strain sensors, providing spatially continuous strain measurement across the blade span. By changing the relative position of the upstream turbine ($WT_1$) to the downstream, waked turbine $WT_2$ across the streamwise and spanwise extent, we map power output, spanwise strain, and accumulated representative fatigue relevant loading across the wake profile. Full wake impingement suppresses blade loading through the associated velocity deficit, while partial wake overlap generates the strongest load intermittency and highest relative fatigue relevant loading, despite an intermediate power recovery. A combined performance-to-loading metric shows this partial-wake regime offers the least favourable trade-off between energy yield and structural loading. These results show that minimising partial-wake exposure, not only mean velocity deficits, should be a design consideration for wind-farm layout and turbine spacing.}

\maketitle

\begin{boxtext}

{\mathversion{bold}\textbf{Impact Statement}}
Wake-induced blade loading has remained difficult to characterise experimentally because conventional strain gauges provide only sparse measurements along the blade span. Here, we combine distributed Rayleigh-backscattering fibre-optic sensing with a controlled two-turbine wind tunnel arrangement to obtain continuous blade strain measurements under systematically varied wake conditions. This enables, for the first time, simultaneous mapping of blade loading, fatigue-relevant dynamics and power production across the wake generated by an upstream turbine.
The measurements reveal that partial wake overlap, rather than complete wake immersion, generates the most severe fatigue-relevant loading despite only modest gains in power production. This finding highlights the importance of considering cyclic transitions between aerodynamic operating states when designing wind-farm layouts and wake-steering strategies. More broadly, the methodology provides an experimental framework for evaluating the structural consequences of wake interactions, supporting future fatigue-aware wind farm optimisation.
\end{boxtext}

\section{Introduction}

Pressure on wind energy to establish itself as a cornerstone of global decarbonisation strategies has accelerated the development of wind energy technology, driving a continuous upscaling of wind turbine size. Modern blades are increasingly thin and flexible \citep{moreno2025_center}, and the newest offshore wind turbines exceed rotor diameters of $300\mathrm{m}$ \citep{website_offshorebiz}, with even larger designs projected \citep{Veers2019Grand}. While this scaling reduces the cost of energy through enhanced harvesting capacity, it also introduces challenges in turbine design, operation, and wind-farm layout optimisation.

Simply enlarging turbine spacing in proportion to rotor diameter is unfeasible: it would produce prohibitively large wind-farm footprints, inflate infrastructure costs, and lead to inefficient use of offshore lease zones. As wind farms become more compact (in a relative sense), downstream machines will operate more frequently in the wakes of their upstream counterparts, experiencing both power deficits and elevated fatigue loads \citep{Thomsen1999Fatigue, Barthelmie2009Modelling, Stevens2017Flow, Moens2022a, Moens2022b, Pacheco2024Experimental}. The near- and far-wake regions of a turbine have been characterised extensively through both experimental and analytical means \citep{Vermeer2003, porteagel2020, Bastankhah2014}, and the associated velocity deficit is known to depend strongly on the operating condition and induction of the upstream rotor \citep{Adaramola2011WakeEffects}. In addition to the mean deficit, wakes meander laterally under the influence of large-scale atmospheric or facility-scale turbulence \citep{Larsen2008WakeMeandering}, further modulating the unsteady inflow experienced by any downstream machine. From a structural standpoint, waked turbines are therefore subject to inflow conditions that differ markedly from the free-stream, experiencing elevated turbulence intensity ($TI$) and modified integral length scales ($\mathcal{L}$) relative to undisturbed conditions \citep{Thomsen1999Fatigue, Pacheco2024Experimental, Hodgson2025, Frandsen2007}. In addition, significant mean velocity gradients can exist across the \enquote{waked} rotor swept area, imposing spatially non-uniform aerodynamic loading across the blades.

Considerable effort has been directed toward optimising wind-farm layout and individual turbine control to maximise power output (see \citep{Fleming2015, Gebraad2016, stevens2017, campagnolo2020} and references therein). However, wind turbine wake-induced structural loading has received comparatively less attention, and the experimental characterisation of blade dynamics under waked inflow conditions remains limited. Where fatigue-relevant loading has been addressed directly, this has largely relied on numerical actuator-disc or aeroelastic simulations \citep{Moens2022a, Moens2022b}, or on full-scale field campaigns constrained to a small number of naturally occurring wake encounters \citep{Pacheco2024Experimental}. Controlled, systematically varied experimental characterisation of blade-level loading across a wide range of \enquote{waked} conditions spanning both lateral offset and streamwise separation remains comparatively scarce. Accurate predictions of how wakes induce fatigue damage, by introducing both velocity deficits and coherent flow dynamics---expected to be more damaging to structures than incoherent motions---to the inflow to downstream turbines, are therefore of considerable practical relevance. In fact, flow structures with increased spatial coherence have been associated with an increased induced structural response \citep{francisco2, maryami}.

Blade load monitoring has relied on discrete point sensors, most commonly strain gauges or fibre Bragg grating (FBG) arrays bonded to a handful of spanwise stations \citep{Schroeder2006}. While such approaches have proven reliable for tracking root bending moments and overall fatigue accumulation, they offer only sparse spatial coverage and can therefore miss localised or spanwise-varying features of the aerodynamic loading, particularly under the spatially non-uniform inflow generated by an upstream wake. The structural response of wind turbine blades to waked inflow is inherently distributed: it depends on both local aerodynamic conditions and global rotor dynamics (see \citet{francisco3} for more details), and cannot be adequately characterised by point measurements alone. Distributed fibre-optic sensing based on Rayleigh backscattering offers a route to overcome this limitation, providing spatially continuous strain measurement along the full length of an optical fibre \citep{francisco1, Li2025Shape} and enabling reconstruction of the strain field with a resolution unattainable with discrete point sensors.

To address these limitations, the present study investigates how wake-induced inflow modifies the blade loading experienced by a downstream wind turbine using spatially distributed fibre-optic sensing. Specifically, the work addresses three questions: (i) how the mean and fluctuating blade dynamics evolve with streamwise and lateral turbine spacing, (ii) which wake configurations produce the greatest fatigue-relevant loading, and (iii) how these competing effects influence the trade-off between power production and relevant load dynamics. To this end, experiments were conducted in a large wind tunnel using two $1\,\mathrm{m}$-diameter wind turbine models, one of which was instrumented with distributed Rayleigh backscattering sensors to provide continuous spanwise strain measurements.

\section{Experimental Methodology}

Experiments were conducted in the large test section of the $10' \times 5'$ wind tunnel at Imperial College London, with a cross section of $5.7 \times 2.8~\mathrm{m}^2$, and $18~\mathrm{m}$ in length. Two horizontal-axis wind turbines of rotor diameter $D = 1.0 \mathrm{m}$ were designed using blade element momentum (BEM) theory, and optimised to operate at a tip-speed ratio $\lambda= 4$ (where $\lambda$ is defined as: $\lambda = 0.5 \Omega D/U_{\infty}$, with $\Omega$ the turbine's rotational speed, in $\mathrm{rad/s}$ ).
The blades employ low Reynolds number profiles along their span (SD2030, SG6043 and SG6051 at $r/R<0.15$, $0.15 \leq r/R\leq 0.75$ and $r/R > 0.75$ respectively, where $R$ corresponds to the radius of the turbine). The rotor and the blades were $3$D printed, and are made of Polylactic acid (PLA).
Further details on the rotor design and turbine operation are given in \citet{francisco3}.
Each turbine is mounted on a vertical support strut, with the rotor axis horizontal and aligned with the incoming flow direction.
The upstream wind turbine is subjected to a constant $U_\infty = 2.8 \mathrm{m/s}$, setting the smallest blade chord-based Reynolds number to $Re_{c}\approx 12,000$, and the diameter-based Reynolds number to $Re_{D}\approx 200,000$.

A set of Irwin spires was installed at the inlet of the upper test section, following the design and characterisation presented by \citet{Nair2025ABL} (specifically, Set 2 (TS)). This configuration yields an approximately spanwise-uniform mean velocity profile across the test section. Although the spires generate a developing boundary layer, its influence is confined to elevations below the lowest blade passage, resulting in negligible mean velocity shear across the rotor swept area \citep{Nair2025ABL}. Consequently, the wind turbine models operate under an essentially uniform inflow.

\begin{figure}
  \raggedright
  \includegraphics[width=\textwidth]{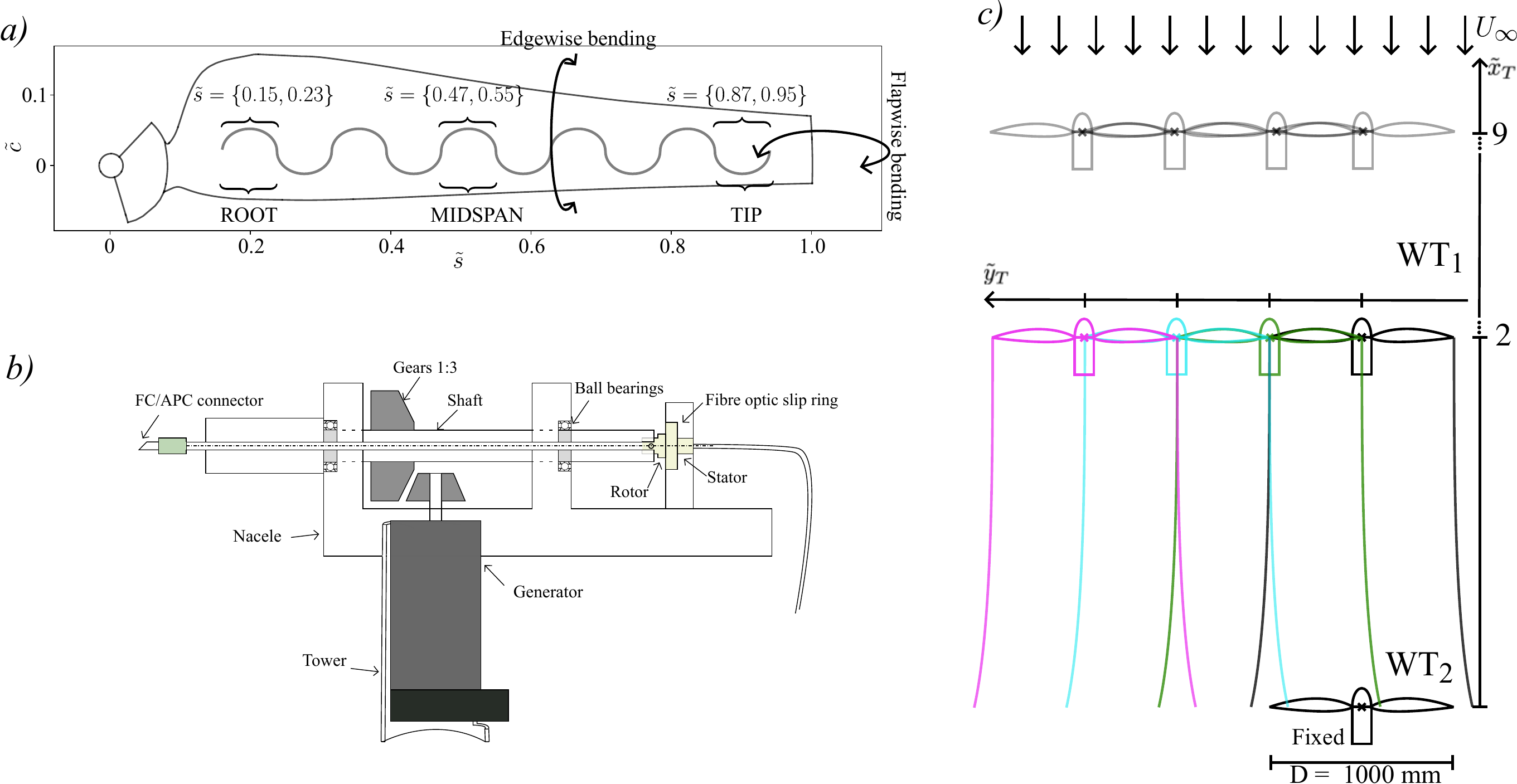}
  \caption{\textit{a)}: Schematic of the fibre-optic sensing layout used, and $3$ regions of interest under analysis - $\mathrm{ROOT}$, $\mathrm{MIDSPAN}$ and $\mathrm{TIP}$. \textit{b)}: Schematic of the wind turbine set-up inside of its nacelle, where the slip ring is set in line with the driving shaft of the turbine's rotor. \textit{c)}: Schematic of representative waked conditions to which the instrumented turbine is subjected.}
  \vspace{-0.6cm}
  \label{fig1}
\end{figure}

One of the turbines is set upstream, referred to as $WT_1$, the other is fixed at a downstream position, referred to as $WT_2$. $WT_2$ is then exposed to \enquote{waked} inflow conditions, generated by the upstream $WT_1$ turbine. 
One of the blades of $WT_2$, the \enquote{waked} turbine, is instrumented with a fibre-optic laid out in a sinusoidal profile on the pressure side of the blade, designed to resolve both spanwise and chordwise dynamics \citep{Li2025Shape,francisco3}. 
The fibre path was defined using a reference grid of pitch $a = 40\,\mathrm{mm}$, producing a sinusoidal path along the blade's span (see figure \ref{fig1} \textit{a)}).
This configuration yields rosettes of both chordwise and spanwise strain across the blade's span. 
The fibre is routed through the hub using an in-line optical slip-ring, ensuring uninterrupted transmission of the optical signal under rotation (detailed view of the setup in figure \ref{fig1} \textit{b)}). More details in \citep{francisco3}.
Data acquisition is performed with LUNA ODiSI-B \citep{francisco1}, providing distributed strain measurements with $\Delta s = 2.6\,\mathrm{mm}$ resolution across the fibre's extent at a fixed acquisition frequency of $f_{\mathrm{acq}} = 100~\mathrm{Hz}$.
Strain data during \enquote{wind-on} conditions was acquired for a total of $T=120~\mathrm{s}$.
Before and after each acquisition, the tunnel was set under \enquote{wind-off} conditions for $T=60~\mathrm{s}$ (generating a quiescent background) and the rotational speed of $WT_2$ was matched to the rotational speed under \enquote{wind-on} conditions. This was done to get an \emph{approximation} of the isolated contribution to the acquired strain from gravitational+centrifugal sources. We highlight \emph{approximation} because even though the atmosphere is quiescent and the expected resulting aerodynamic loading was small, the blades still generate lift as they rotate. Accordingly, a complete isolation of the contribution of gravitational and centrifugal forces to the strain is unattainable, without conducting these tests in a vacuum chamber.
No drift of the strain signal during the tests was observed.

The rotor speed of the two turbines was regulated using a MAXON motor and controller, with the motor operated as a generator (following the methodology of \citep{majid2017}).
The motor is coupled via a $1:3$ bevel-gear set to the shaft of the rotor, allowing control of $\lambda$ during experiments while routing the fibre-optic slip ring in-line with the rotor. 
Figure \ref{fig1} \textit{c)} presents a schematic of the conducted experiments, and respective positioning of the two turbines. 
For each test case, turbine $WT_1$ is moved along the wind tunnel's available span and streamwise extent (respectively $\tilde{y}_T = y/D$ and $\tilde{x}_T=x/D$) to produce different inflow conditions to which $WT_2$ is exposed. The subscript $PT$ denotes the location of $WT_1$ for the different tests conducted.
$WT_1$ was run at two different operating $\lambda_1\in\{4,6.5\}$, generating for each spatial location two different wake profiles. 
$WT_1$ was deliberately chosen to operate at $\lambda_1\in\{4,6.5\}$ to generate a wake velocity profile characteristic of optimum, and above-optimum conditions, representative of two different thrust coefficients ($C_{T,1}$, where the thrust coefficient $C_T = \frac{8 T}{\rho \pi D^2 U_{\infty}^2}$). In fact, $C_T$ has been thoroughly linked to near wake kinematics and is a fundamental parameter in one of the most famous analytical wind turbine wake models \citep{Bastankhah2014}.
Considering the origin of the Cartesian axis at the centre of $WT_2$, $WT_1$ was positioned such that a total of $20$ different inflow \enquote{flavours} were generated for $WT_2$, spanning $\tilde{x}_T\in\{2,3,4,5,7,9\}$ and $\tilde{y}_T\in\{0.0,0.5,1.0,1.5\}$ (notice that the positive $\tilde{x}_T$ direction adopted here is opposite to the streamwise flow direction).
$WT_2$ was operated at design conditions for all cases (\textit{i.e.}, $\lambda_2=\lambda_d=4$). $WT_1$ was moved carefully, but manually across the span of the wind tunnel. This potentially introduced misplacement errors across the tunnel's span. However, we estimate these to be relatively small when compared to the diameter of the wind turbine models, of order $\delta_{x,y}/D \approx\pm5\%$ (where $\delta_{x,y}$ corresponds to the deviation from the desired location for $WT_1$ to its true position).

The current output ($\mathrm{I_{out}}$) from the generator connected to $WT_2$ was acquired during operation using the internal diagnostics module of the MAXON ESCON 50/5 controller. The generator terminal voltage ($\mathrm{V_{out}}$) was estimated from the measured rotational speed ($\Omega$) using the motor's speed constant ($\kappa = 116~\mathrm{rpm,V^{-1}}$) provided by the manufacturer, according to $\mathrm{V_{out}}=\frac{\Omega}{\kappa}$, where $\Omega$ was obtained from the encoder mounted inline with the generator. The electrical power produced by $WT_2$ was then estimated as $P_2=\mathrm{I_{out}}\times\mathrm{V_{out}}$.
The uncertainty associated with the estimated electrical power is determined by the uncertainties in the current and encoder-based rotational speed measurements. Following the manufacturer's guidance, the encoder accuracy was assumed to be within $\pm1~\mathrm{rpm}$ \citep{maxonspeedacc}, corresponding to an absolute voltage uncertainty of $\pm1/\kappa \approx \pm0.0086~\mathrm{V}$. As the manufacturer does not provide an accuracy specification for the internally reported current measurement, a conservative uncertainty of $\pm1\%$ of the measured current was assumed. The relative uncertainty associated with the estimated electrical power was then obtained using standard uncertainty propagation, $\frac{\epsilon_P}{P} = \sqrt{\left(\frac{\epsilon_I}{I}\right)^2+\left(\frac{\epsilon_\Omega}{\Omega}\right)^2}$, where $\epsilon_I$ and $\epsilon_\Omega$ denote the uncertainties associated with the current and rotational speed measurements, respectively. Based on these assumptions, the maximum relative uncertainty in the estimated electrical power over all operating conditions was found to be approximately $1.05\%$ ($\approx 0.0464\,\mathrm{W}$). Finally, note that the electrical measurements were acquired downstream of the gearbox and mechanical transmission components. Consequently, any mechanical transmission losses between the rotor and the generator are inherently accounted for in the reported power estimates.

\begin{figure}
	\centering
	\raisebox{1.8in}{\textit{a)}}\includegraphics[width=.8\textwidth]{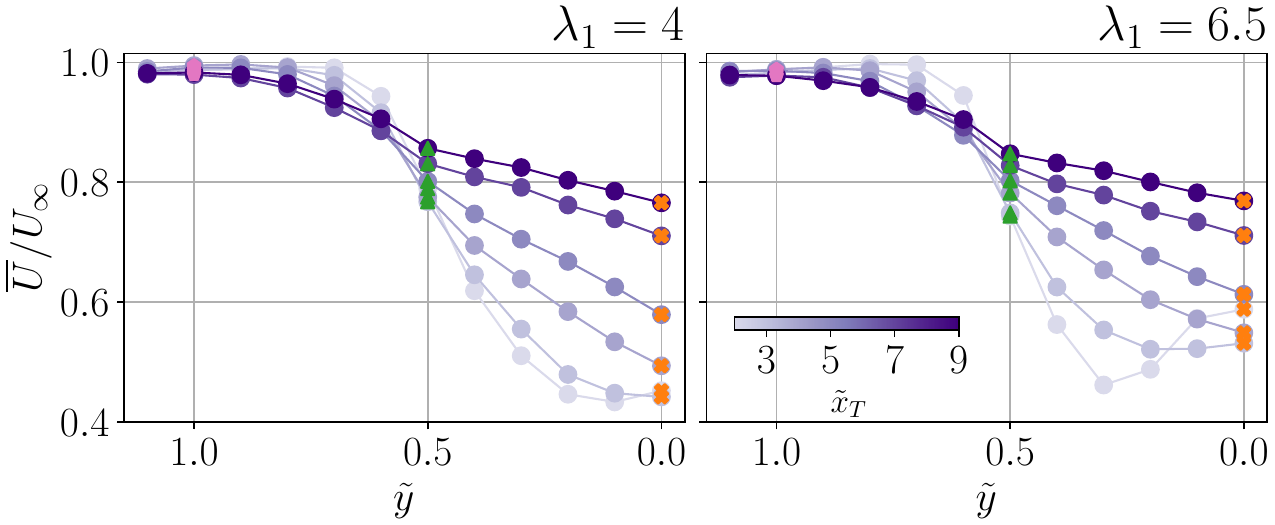}
	\raisebox{1.8in}{\textit{b)}}\includegraphics[width=.8\textwidth]{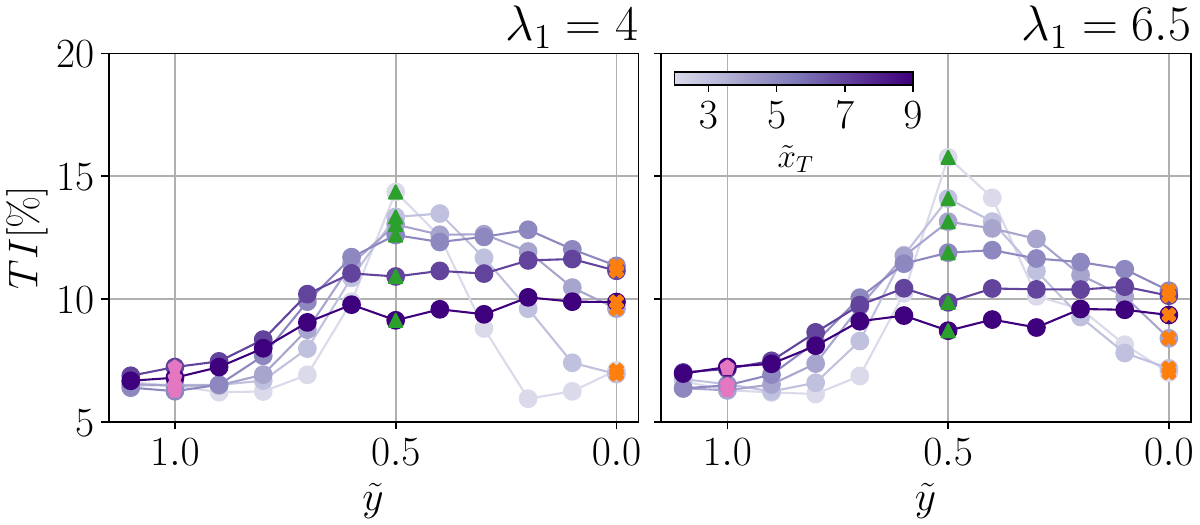}
	\caption{Normalised time-averaged velocity profiles $\overline{U}/U_{\infty}$ \textit{a)} and $TI$ \textit{b)} inflow profiles to which $WT_2$ is subjected, spanning from $\tilde{y}\in\{0,1.1\}$.}
	\label{fig:wt3_fig2}
\end{figure}

Preliminary hot-wire data was obtained to assess the inflow conditions to which $WT_2$ is exposed during the \enquote{waked} tests, to be able to fine-tune the relative operating rotational speed of $WT_2$ ($\Omega_2$), to be able to set $\lambda_2=\lambda_d=4$ considering the centerline inflow mean velocity at hub-height.
The hot-wire data was taken with a Dantec Streamline Pro system acquiring at a sampling frequency of $10,000 \mathrm{Hz}$.
A rack of $6$ single-component hot wire probes was positioned in the same place as $WT_2$ during the \enquote{waked} tests, and $WT_1$ was moved with respect to the fixed hot-wire rack, spanning $2\leq\tilde{x}_T\leq9$ and $0\leq\tilde{y}_T\leq1.1$, aiming to cover the full lateral extent over which the mean velocity and $TI$ profiles depart from free-stream conditions (see figure~\ref{fig:wt3_fig2}). At $\tilde{y}_T=1.5$, the inflow was assumed to be undisturbed by the upstream wake and $WT_2$ was exposed to close to free-stream conditions.
Panel \textit{a)} from figure \ref{fig:wt3_fig2} presents the normalised mean velocity profiles acquired ($\overline{U}/U_{\infty}(\tilde{x}_T, \tilde{y})$), and panel \textit{b)} the resulting turbulence intensity profiles ($TI = \sqrt{\overline{u^{\prime~2}}}/\overline{U}$), where $\overline{\cdot}$ and $u^{\prime}$ correspond respectively to the time-averaged operator, and the velocity fluctuations obtained through Reynolds decomposition ($u^{\prime} = U - \overline{U}$).
The position of the hub-centre of $WT_2$ with respect to the velocity profiles at $\tilde{y}_T\in\{0,0.5,1.0\}$ for the different $\tilde{x}_T$ is highlighted with, respectively, orange, green and pink markers.

The half-wake time-averaged velocity and $TI$ profiles follow the nominal trend, presenting a Gaussian-like deficit profile, with the velocity decaying smoothly from $\overline{U}/U_{\infty}\approx1$ at the outer edge ($\tilde{y}\approx1$) towards the rotor centre, and a strong shear emerging at $\tilde{y}\approx0.5$ at the tip location of the blades, especially pronounced for streamwise locations close to the turbine.
The $TI$ profiles corroborate this picture, exhibiting a pronounced peak near $\tilde{y}\approx0.5$ with the exception of the cases in which $WT_1$ is placed farthest upstream, coinciding with the tip vortex region and the high-shear zone of the velocity profiles. Away from the shear layer, $TI$ decayed towards the free-stream background value for $\tilde{y}\gtrsim0.8$. With increasing downstream distance $\tilde{x}_T$, the $TI$ peak broadens and attenuates as the tip vortices break down and turbulence diffuses laterally. The effect of $\lambda$ is also evident, especially at $\tilde{x}_T<5$: the $\lambda_1=6.5$ case exhibits both a deeper velocity deficit and elevated $TI$ levels, consistent with the decreased porosity across the rotor plane.

\section{Operational conditions}

\begin{figure}
	\centering
	\includegraphics[width=.85\textwidth]{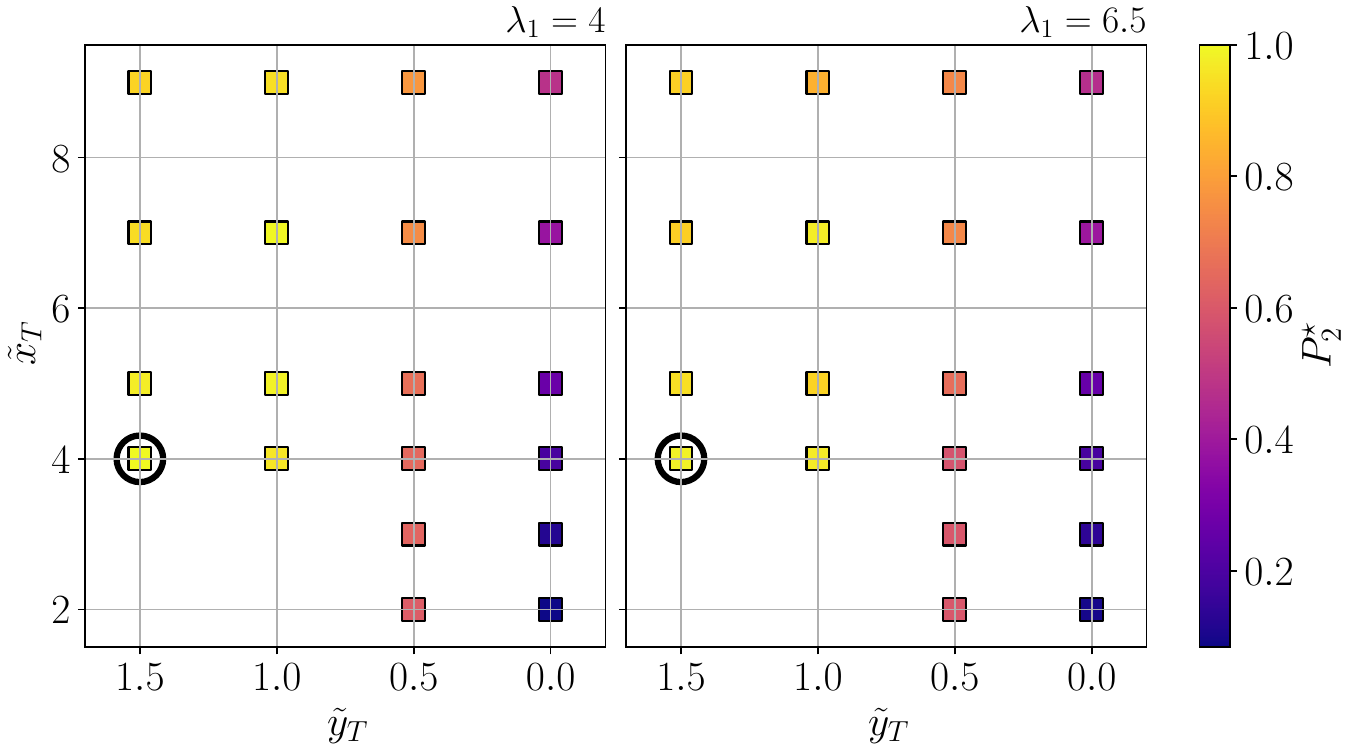}
	\caption{Normalised power $P_2^{\star}$ produced by $WT_2$ at each inflow condition, as a function of the lateral and streamwise position of $WT_1$ relative to $WT_2$, for $\lambda_1 = 4$ (left) and $\lambda_1 = 6.5$ (right). The circled marker denotes the position in which the turbine produced most power, experiencing the inflow closest to undisturbed free-stream conditions.}
	\label{fig:wt3_fig3}
\end{figure}

Figure~\ref{fig:wt3_fig3} presents the normalised power output $P_2^\star$ of the waked turbine $WT_2$ across the full experimental matrix, for both upstream tip-speed ratios $\lambda_1 \in \{4, 6.5\}$.
The normalisation is performed with respect to the unwaked power output of $WT_2$, such that $P_2^\star = 1$ corresponds to the normalised power output from conditions equivalent to clean-free-stream operation (at $(\tilde{x}_T=4, \tilde{y}_T=1.5)$).
 
A clear and consistent dependence on both $\tilde{y}_T$ and $\tilde{x}_T$ is observed.
At large lateral offsets ($\tilde{y}_T = 1.5$), $P_2^\star \approx 1$ for all streamwise separations and both tip-speed ratios, confirming that $WT_2$ operates effectively outside the wake of $WT_1$ (or very close to it) at this lateral position.
As $\tilde{y}_T$ decreases towards full wake impingement ($\tilde{y}_T = 0$), $P_2^\star$ decreases substantially, with the strongest deficits observed at the smallest streamwise separations.
At $\tilde{x}_T = 2$, $P_2^\star$ drops to approximately $0.15$--$0.2$ at full wake overlap ($\tilde{y}_T = 0$), reflecting the large velocity deficit in the near-wake region.
This deficit recovers progressively with increasing $\tilde{x}_T$ across the range of tested $\tilde{y}_T$, consistent with the wake velocity recovery observed in figure~\ref{fig:wt3_fig2}.
Comparing $P_2^{\star}$ for the two $\lambda_1$ tested, the power deficit maps are broadly similar in structure, following similar time-averaged velocity profiles at the near-wake of $WT_1$ produced by the two considered operating conditions ($\lambda_1$).
 
\section{Time-averaged blade loading}
 
\begin{figure}
	\centering
	\raisebox{2.8in}{\textit{a)}}\includegraphics[width=.88\textwidth]{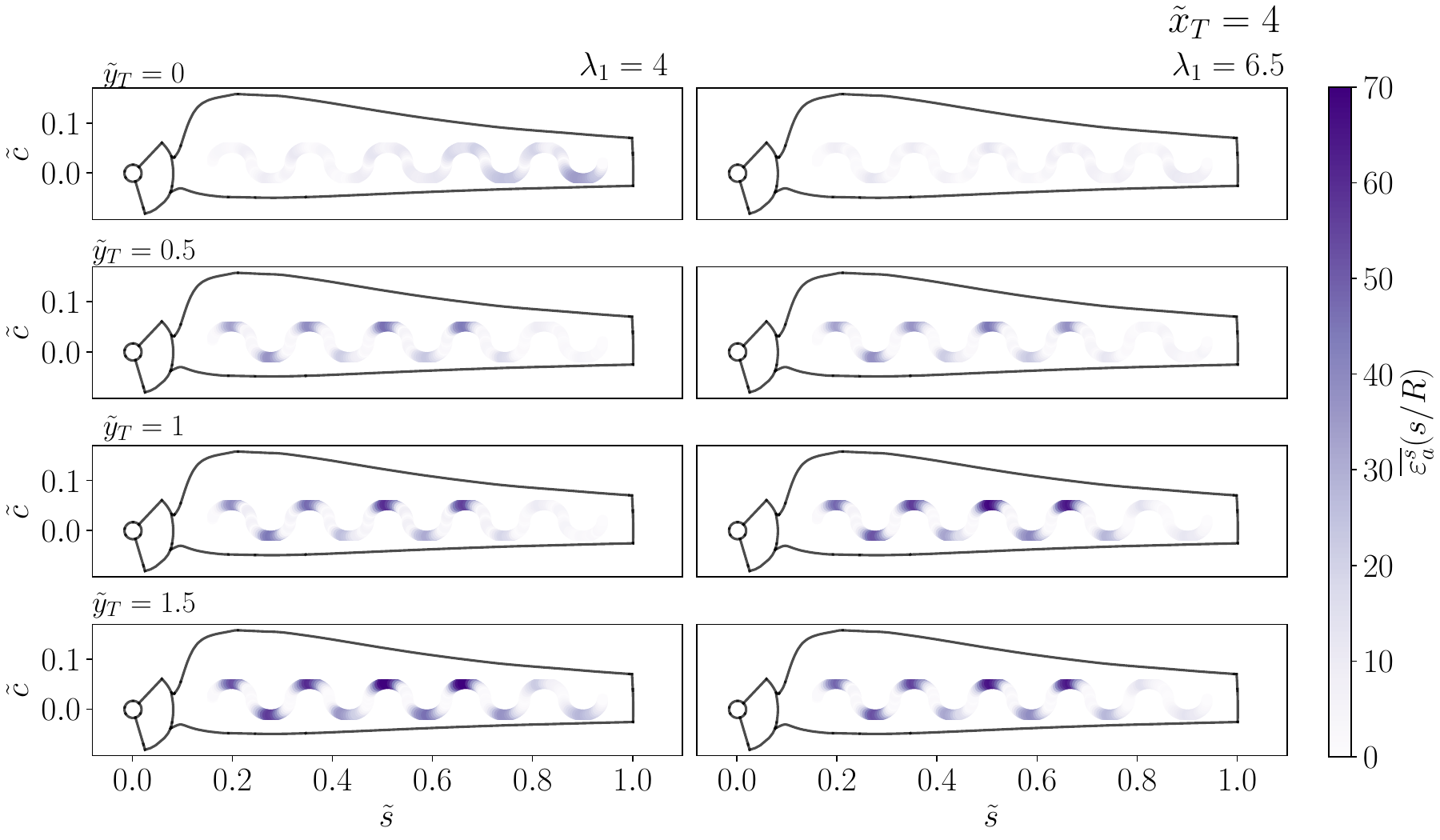}
	\raisebox{2.80in}{\textit{b)}}\includegraphics[width=.88\textwidth]{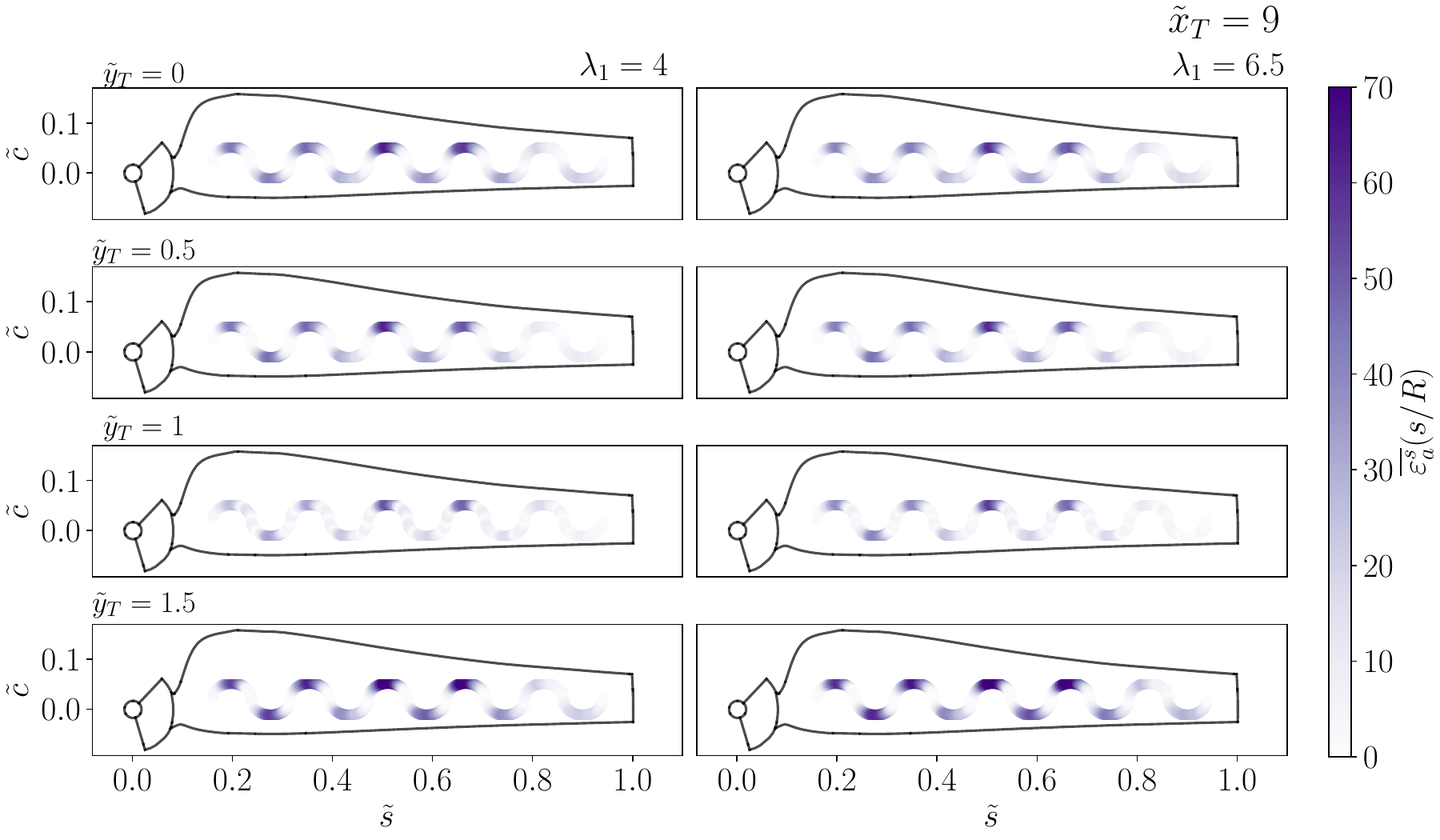}
	\caption{Time-averaged spanwise strain $\overline{\varepsilon}_a^s(\tilde{s})$ distributed along the blade span for \textit{a)} $\tilde{x}_T = 4$ and \textit{b)} $\tilde{x}_T = 9$, across lateral positions $\tilde{y}_T \in \{0, 0.5, 1.0, 1.5\}$ (rows) and both tip-speed ratios $\lambda_1 \in \{4, 6.5\}$ (columns).}
	\label{fig:wt3_fig4}
\end{figure}

The time-averaged aerodynamic-driven spanwise strain distributions $\overline{\varepsilon}_a^s(\tilde{s})$ are presented in figures~\ref{fig:wt3_fig4}\textit{a)} and \ref{fig:wt3_fig4}\textit{b)} for two representative streamwise positions, $\tilde{x}_T = 4$ and $\tilde{x}_T = 9$, respectively.
 
At $\tilde{x}_T = 4$, a strong dependence of the strain distribution on the lateral wake position $\tilde{y}_T$ is evident for both $\lambda_1$ tested.
Under full wake impingement ($\tilde{y}_T = 0$), $\overline{\varepsilon}_a^s(\tilde{s})$ is markedly reduced across the entire blade span compared to partially-waked and unwaked conditions.
This is consistent with the velocity deficit experienced by $WT_2$ at close spacings, as the reduced inflow momentum directly suppresses the magnitude of the aerodynamic loading.
Consistent with the power measurements, the mean strain recovers with increasing streamwise spacing as the wake deficit weakens.
At half-waked conditions ($\tilde{y}_T=0.5$), the rotor is exposed to increased velocity when compared to fully-immersed waked conditions, and the strain distribution recovers partially, with measurable loading observed across the midspan and tip regions for both streamwise stations. 
The condition of highest mean strain magnitude at $\tilde{x}_T = 4$ is observed for $\tilde{y}_T = 1.0$--$1.5$, where $WT_2$ operates predominantly outside the wake, yielding strain levels approaching the free-stream baseline; as shown in section~\ref{sec:fatigue}, this is distinct from the condition of highest fatigue-relevant damage, which instead occurs under partial wake exposure.

\begin{figure}
	\centering
	\raisebox{1in}{\textit{a)}}\includegraphics[width=0.7\textwidth]{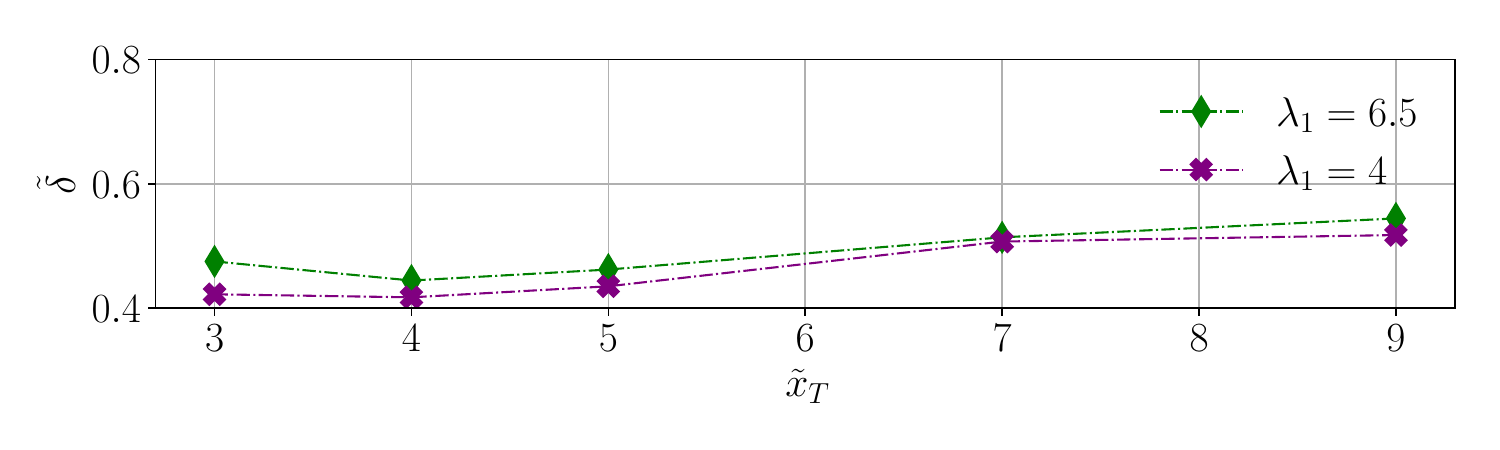}
	\raisebox{1in}{\textit{b)}}\includegraphics[width=0.7\textwidth]{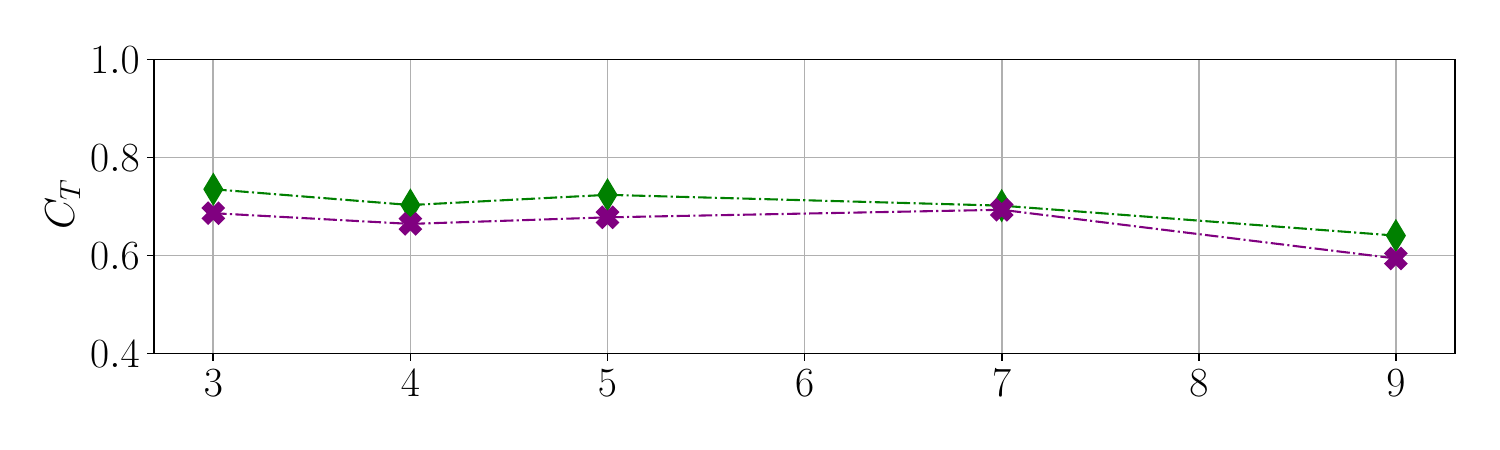}
		\caption{\textit{a)} Thrust coefficient, $(C_{T,1})$, estimated from the wake momentum deficit measured for each upstream position of WT1. \textit{b)} Diameter-normalised wake half-width, $(\tilde{\delta})$, obtained from Gaussian fits to the velocity-deficit profiles.}
	\label{fig:wt3_cthw}
\end{figure}

Virtually no differences in the time-averaged aerodynamic distributions are observed between the two tested tip-speed ratios. Given that the two $\lambda_1$ produce comparable mean wake velocity profiles (figure~\ref{fig:wt3_fig2}), this similarity in mean strain is expected. To further quantify the aerodynamic loading and wake development, the thrust coefficient, $C_T$, and the diameter-normalised wake half-width, $\tilde{\delta}$ generated by $WT_1$, were estimated from the mean velocity profiles at all surveyed streamwise stations, though the resulting estimates are most reliable for $\tilde{x}_T\geq3$, where pressure recovery is expected to be largely complete and the wake evolves towards a smoother, more axisymmetric deficit profile - see figure \ref{fig:wt3_fig2} \textit{a)}. The thrust coefficient was obtained from an axisymmetric momentum-deficit formulation, using a standard atmospheric air density $\rho = 1.225~\mathrm{kg/m^3}$, while $\tilde{\delta}$ was determined by fitting a Gaussian distribution to the normalised velocity deficit and extracting the half-deficit width. The resulting distributions are shown in figure~\ref{fig:wt3_cthw}.

The estimated thrust coefficients confirm the proximity of the two operating conditions: across all tested spacings both yield $C_T \in [0.63,0.74]$, with $\Delta C_T \approx 0.03$–$0.05$ throughout. Despite this modest difference, $\lambda_1 = 6.5$ produces a systematically wider wake, with $\tilde{\delta}\approx0.51$ compared with $\approx0.43$ for $\lambda_1 = 4$ at $\tilde{x}_T=2$, indicating that the extracted momentum is distributed over a larger radial extent. The combination of a slightly larger $C_T$ and a weaker centreline velocity deficit is consistent with a redistribution of the extracted momentum into wake rotation and radial spreading, rather than exclusively into axial flow deceleration.

As a result, the axial velocity deficit alone does not fully reflect the total momentum extraction from the flow \citep{FuentesNoriega2025SwirlingWake}.
Since the present measurements rely on single-component hot-wire probes, only the streamwise velocity component is captured, and the contribution of tangential momentum remains unresolved. Consequently, differences in wake structure between the two operating conditions may be underestimated, particularly in regimes where swirl plays a significant role.

\begin{figure}
	\centering
	\includegraphics[width=\textwidth]{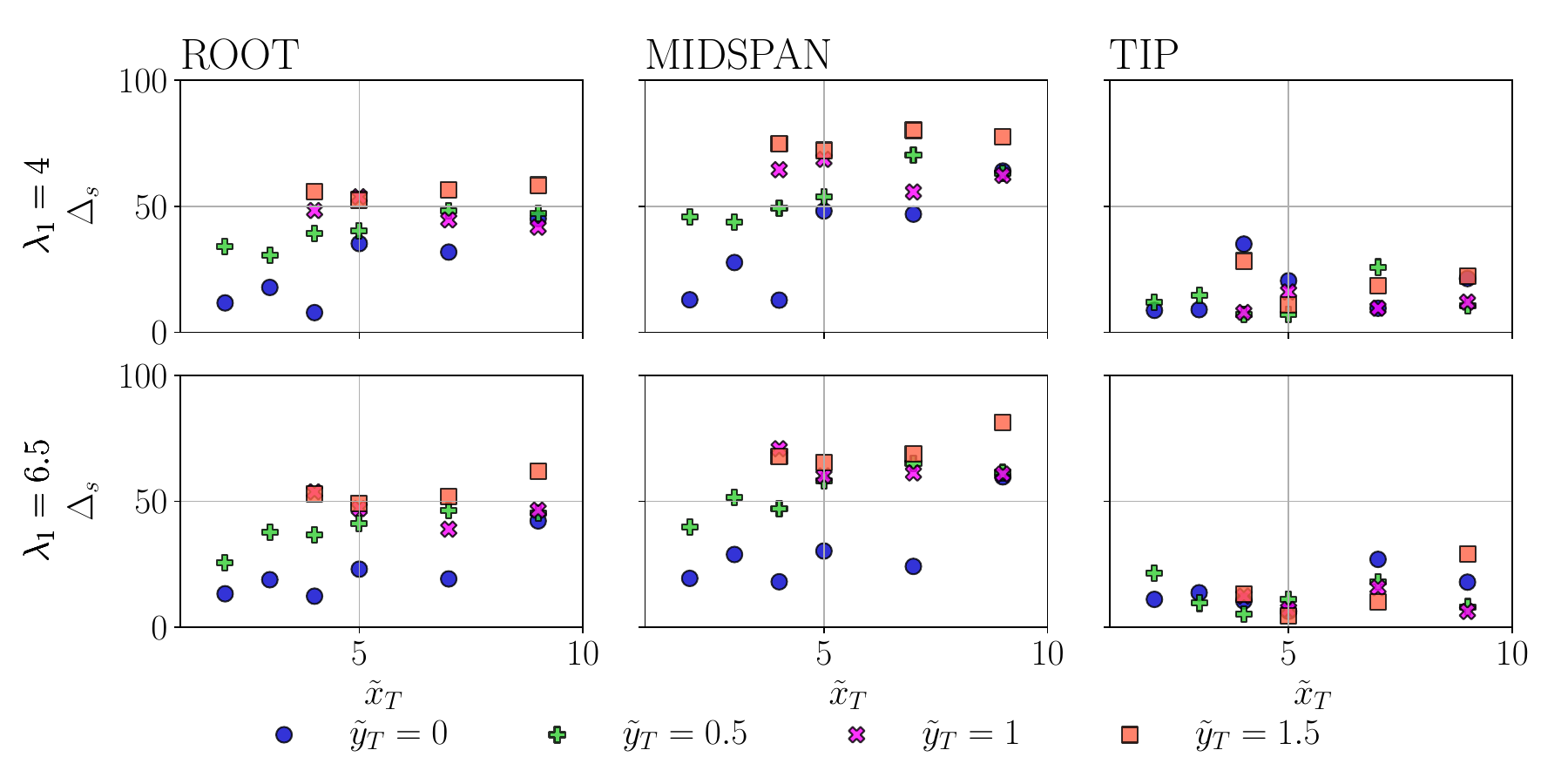}
	\caption{Spanwise bending proxy $\Delta_s$ evaluated at the $\mathrm{ROOT}$ (left), $\mathrm{MIDSPAN}$ (centre) and $\mathrm{TIP}$ (right) blade regions, as a function of streamwise separation $\tilde{x}_T$, for lateral positions $\tilde{y}_T \in \{0, 0.5, 1.0, 1.5\}$ and both tip-speed ratios $\lambda_1 = 4$ (top row) and $\lambda_1 = 6.5$ (bottom row).}
	\label{fig:wt3_fig5}
\end{figure}

Similarly to the analysis of \citep{francisco3}, we assess $\Delta_s$ as a proxy of the time-averaged spanwise loading across operating conditions. $\Delta_s$, defined as the local peak-to-trough difference of $\overline{\varepsilon_{a}^{s}}$ within each region (i.e. $\Delta_s = \mathrm{max}(\overline{\varepsilon_a^s}) - \mathrm{min}(\overline{\varepsilon_a^s})$) provides information on approximate local bending of the blade at each section.
Figure~\ref{fig:wt3_fig5} presents $\Delta_s$ evaluated over the $\mathrm{ROOT}$, $\mathrm{MIDSPAN}$, and $\mathrm{TIP}$, where the difference between the extrema of time-averaged aerodynamic strain is taken over the spanwise extent of the region.

Across all regions and both tip-speed ratios, $\Delta_s$ generally increases with $\tilde{y}_T$ at fixed $\tilde{x}_T$, directly reflecting the inflow conditions: the greater the wake overlap, the smaller is the experienced time-averaged aerodynamic induced strain. The $\mathrm{MIDSPAN}$ carries the largest values and largest spread of strain magnitudes across the span of $\tilde{y}_T$ tested, while the $\mathrm{TIP}$ presents the smallest absolute values overall, consistent with: a reduced chord and local lift contribution; and a smaller accumulated bending moment (relatively to the rest of the blade) as the outboard portion of the blade contributing to the locally experienced bending moment is smaller.

and with the smaller experienced aerodynamic induced moment. For $\tilde{y}_T = 1.5$, the two tip-speed ratios converge to similar $\Delta_s$ levels across all regions, confirming that $WT_2$ operates effectively in free-stream conditions at this lateral offset. Note that the measured strain at a given spanwise location reflects the integrated bending response to the aerodynamic loading acting over the blade outboard of that location, rather than the local load itself. Accordingly, $\mathrm{ROOT}$, $\mathrm{MIDSPAN}$ and $\mathrm{TIP}$ refer only to the measurement locations. Any subsequent association between a sectional strain signal and a particular flow mechanism should therefore be interpreted as qualitative rather than as a direct measurement of the local aerodynamic forcing.

\section{Fluctuating blade strain dynamics}
\begin{figure}
	\centering
	\includegraphics[width=\textwidth]{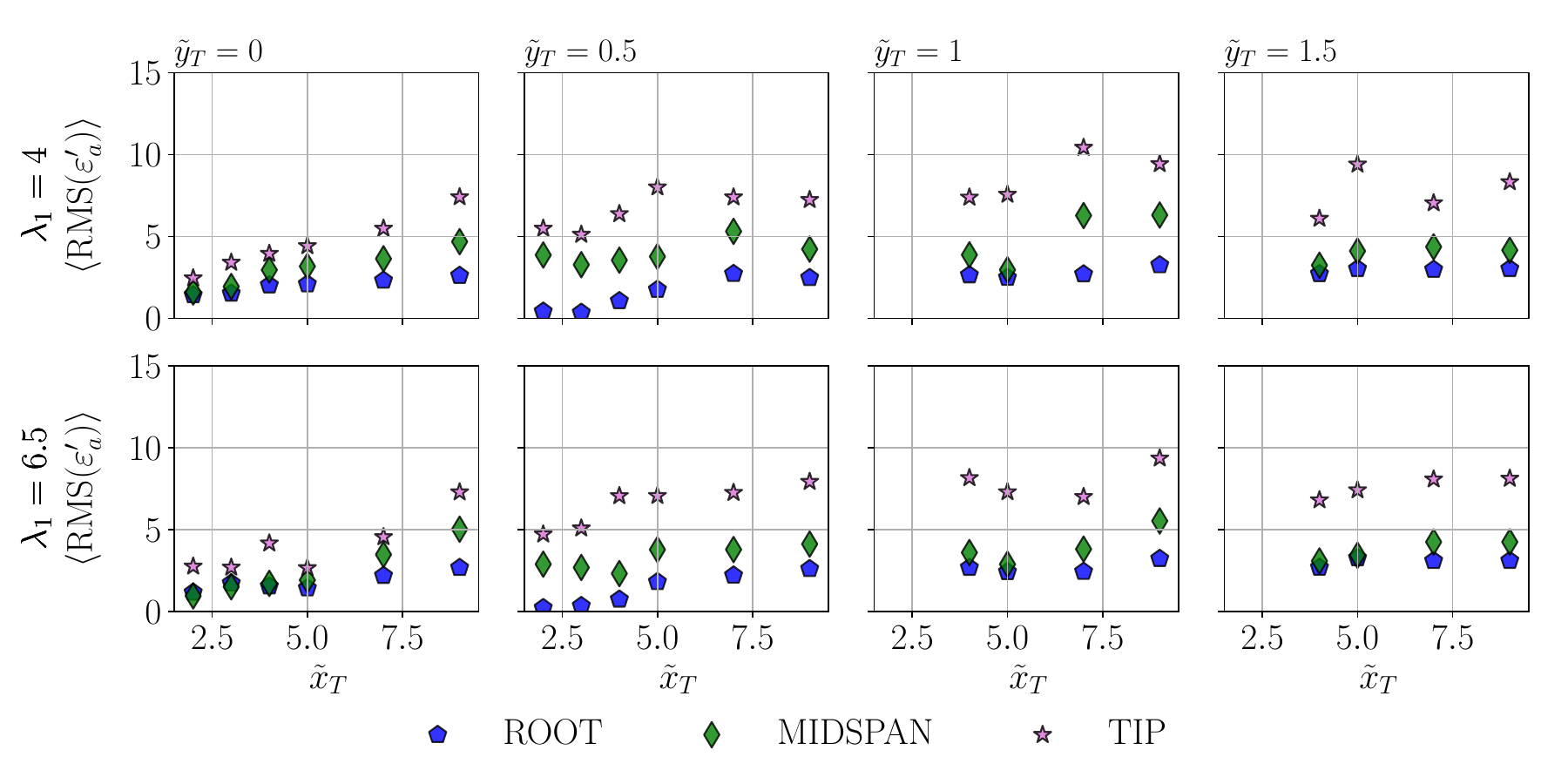}
	\caption{RMS of spanwise strain fluctuations $\gamma \equiv \langle\mathrm{RMS}(\varepsilon_a^\prime)\rangle$ at the $\mathrm{ROOT}$, $\mathrm{MIDSPAN}$ and $\mathrm{TIP}$ blade regions as a function of streamwise separation $\tilde{x}_T$, for lateral positions $\tilde{y}_T \in \{0, 0.5, 1.0, 1.5\}$ (columns) and both tip-speed ratios $\lambda_1 = 4$ (top row) and $\lambda_1 = 6.5$ (bottom row).}
	\label{fig:wt3_fig6}
\end{figure}

The distributions of $\langle\mathrm{RMS}(\varepsilon_a^\prime)\rangle$ as a function of $\tilde{x}_T$, $\tilde{y}_T$, estimated as described in \citep{francisco3}, blade section and $\lambda_1$ are presented in figure~\ref{fig:wt3_fig6}.

The dependence of $\langle\mathrm{RMS}(\varepsilon_a^\prime)\rangle$ on $\tilde{y}_T$ is most pronounced at the $\mathrm{TIP}$ and $\mathrm{MIDSPAN}$. At the $\mathrm{TIP}$, this likely results from a smaller thickness profile when compared to the $\mathrm{ROOT}$, at the $\mathrm{MIDSPAN}$ from the integration of the structural dynamics experienced across the outboard portion of the blade.
Across both tip-speed ratios, the lowest fluctuation levels are consistently observed under fully waked conditions ($\tilde{y}_T=0$), where the reduced inflow velocity suppresses the aerodynamic response. As the turbine moves away from the wake centreline, $\langle\mathrm{RMS}(\varepsilon_a^\prime)\rangle$ increases significantly, indicating a recovery of unsteady loading as the inflow velocity increases. For $\tilde{y}_T \geq 0.5$, however, the fluctuation levels tend to plateau, with similar magnitudes observed for $\tilde{y}_T=1$ and $\tilde{y}_T=1.5$, suggesting that the turbine is effectively operating under near clean free-stream conditions (\textit{i.e.} negligible wake influence).
In the present configuration, despite operating under optimum conditions, the increase in $\langle\mathrm{RMS}(\varepsilon_a^\prime)\rangle$ with $\tilde{y}_T$ can therefore be primarily attributed to the recovery of inflow velocity, which increases the magnitude of the experienced aerodynamic load, rather than to a distinct amplification of strain dynamics within the wake shear layer. An exception is the partially waked case ($\tilde{y}_T=0.5$), where the blades alternately traverse the wake and the free stream during each revolution. In this configuration, the mean strain represents an average between these two loading states, so the calculated RMS also includes the deterministic modulation associated with this periodic switching, in addition to the genuinely unsteady aerodynamic fluctuations.

The streamwise evolution of $\langle\mathrm{RMS}(\varepsilon_a^\prime)\rangle$ reveals distinct behaviours for fully and partially waked conditions.
For $\tilde{y}_T=0$, the fluctuation levels increase monotonically with $\tilde{x}_T$, consistent with the progressive recovery of the wake and the associated increase in aerodynamic loading.
In contrast, for $\tilde{y}_T = 0.5$, $\langle\mathrm{RMS}(\varepsilon_a^\prime)\rangle$ exhibits an initial increase with $\tilde{x}_T$, followed by a clear tendency to plateau beyond $\tilde{x}_T \approx 4$-$5$, with minor variations depending on $\lambda_1$. This reflects the geometric arrangement of $WT_2$ relative to the wake: at small streamwise separations the rotor sits within the high-gradient shear layer, but as the wake widens and diffuses laterally with $\tilde{x}_T$, the local velocity gradient across the rotor disc weakens and the rotor progressively exits this high-shear region.
This interpretation should also be viewed in the context of the intermittent wake boundary (see \citet{Neunaber2020} for more details on the \emph{intermittency ring} surrounding the wake of a wind turbine). The outer region of a wind turbine wake is characterised by alternating incursions of wake and free-stream fluid, such that a blade positioned near the wake edge periodically experiences both flow states. 
For $\tilde{y}_T \geq 1$, the geometric arrangement is reversed: $WT_2$ initially operates in near free-stream conditions and is progressively engulfed by the laterally expanding wake as $\tilde{x}_T$ grows. This explains the gradual increase of $\langle\mathrm{RMS}(\varepsilon_a^\prime)\rangle$ with $\tilde{x}_T$, initially experienced at the $\mathrm{TIP}$ as it emerges firstly onto the higher-velocity region of the flow, whilst being evident at the $\mathrm{ROOT}$ and $\mathrm{MIDSPAN}$ due to the load integration being conducted through bending moments. The convergence of the partial-wake plateau ($\tilde{y}_T=0.5$, $\tilde{x}_T>5$) with the values measured at $\tilde{y}_T=1$ and $1.5$ is therefore a natural consequence of this lateral spreading: in all three cases, $WT_2$ is effectively sampling the diffuse outer edge of the recovering wake.

The increase in $\langle\mathrm{RMS}(\varepsilon_a^\prime)\rangle$ with wake recovery is governed primarily by the increase in mean inflow velocity.
$\tilde{y}_T=0.5$ is the spanwise location richest in $TI$, and does not present a distinct $\langle\mathrm{RMS}(\varepsilon_a^\prime)\rangle$ in comparison to the rest of the cases, strengthening the link between increased mean inflow velocity, and an increased $\langle\mathrm{RMS}(\varepsilon_a^\prime)\rangle$.
As the mean velocity recovers, $WT_2$ operates progressively closer to its design aerodynamic state: the blade sections approach peak $L/D$ (lift to drag ratio), where lift production is most efficient and the unsteady aerodynamic forcing of the structure is maximised. This is the same mechanism as we identified in \citep{francisco3}, where strain fluctuations peaked near $\lambda_d$ regardless of the tested free-stream $TI$. 
These observations motivate a more detailed cycle-based analysis of the loading, such as rainflow counting, to better characterise the fatigue-relevant load dynamics under varying wake conditions.

\section{Relative fatigue induced dynamics}\label{sec:fatigue}

\begin{figure}
	\centering
	\raisebox{2.6in}{\textit{a)}}\includegraphics[width=\textwidth]{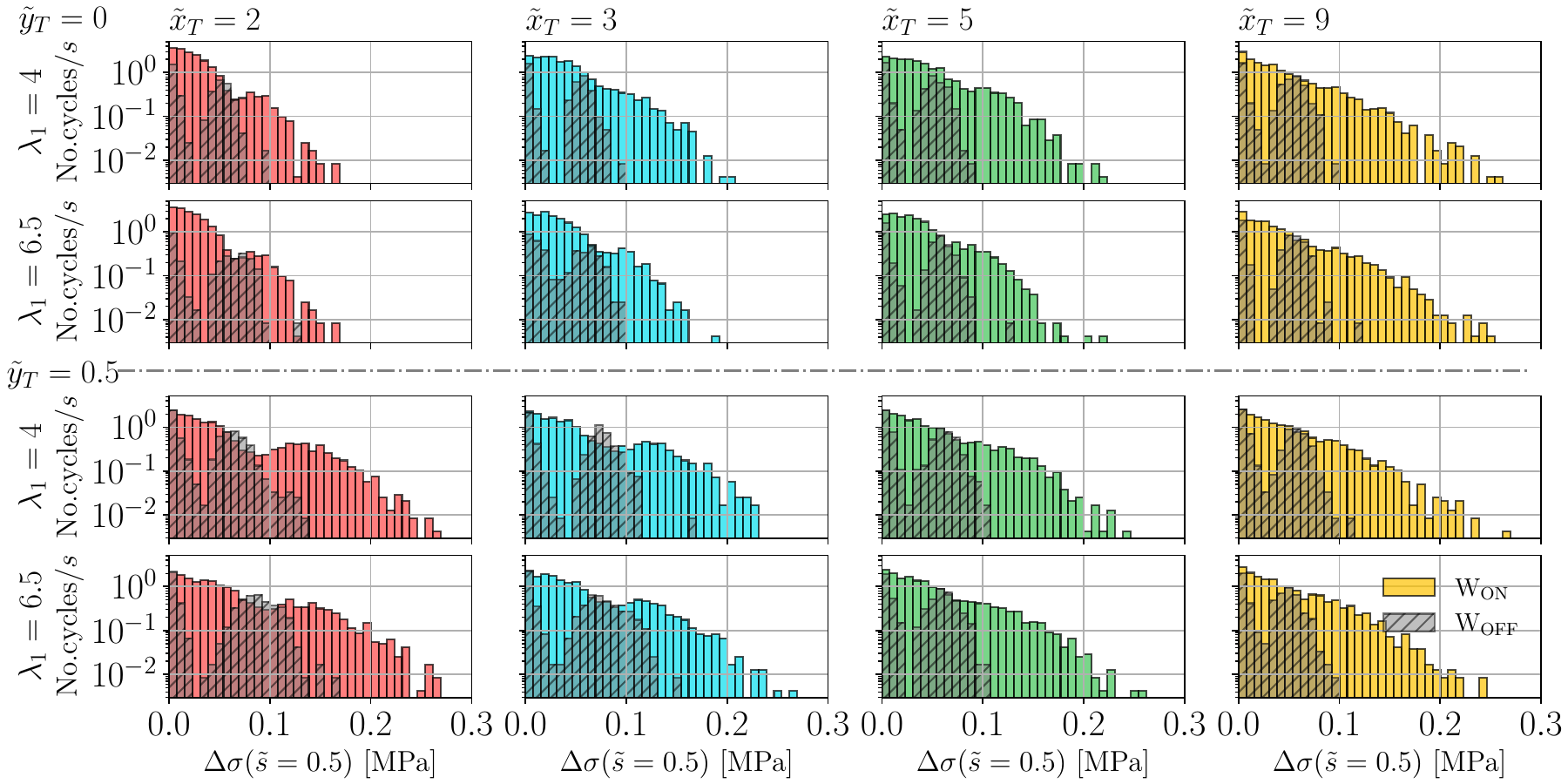}
	\raisebox{2.6in}{\textit{b)}}\includegraphics[width=\textwidth]{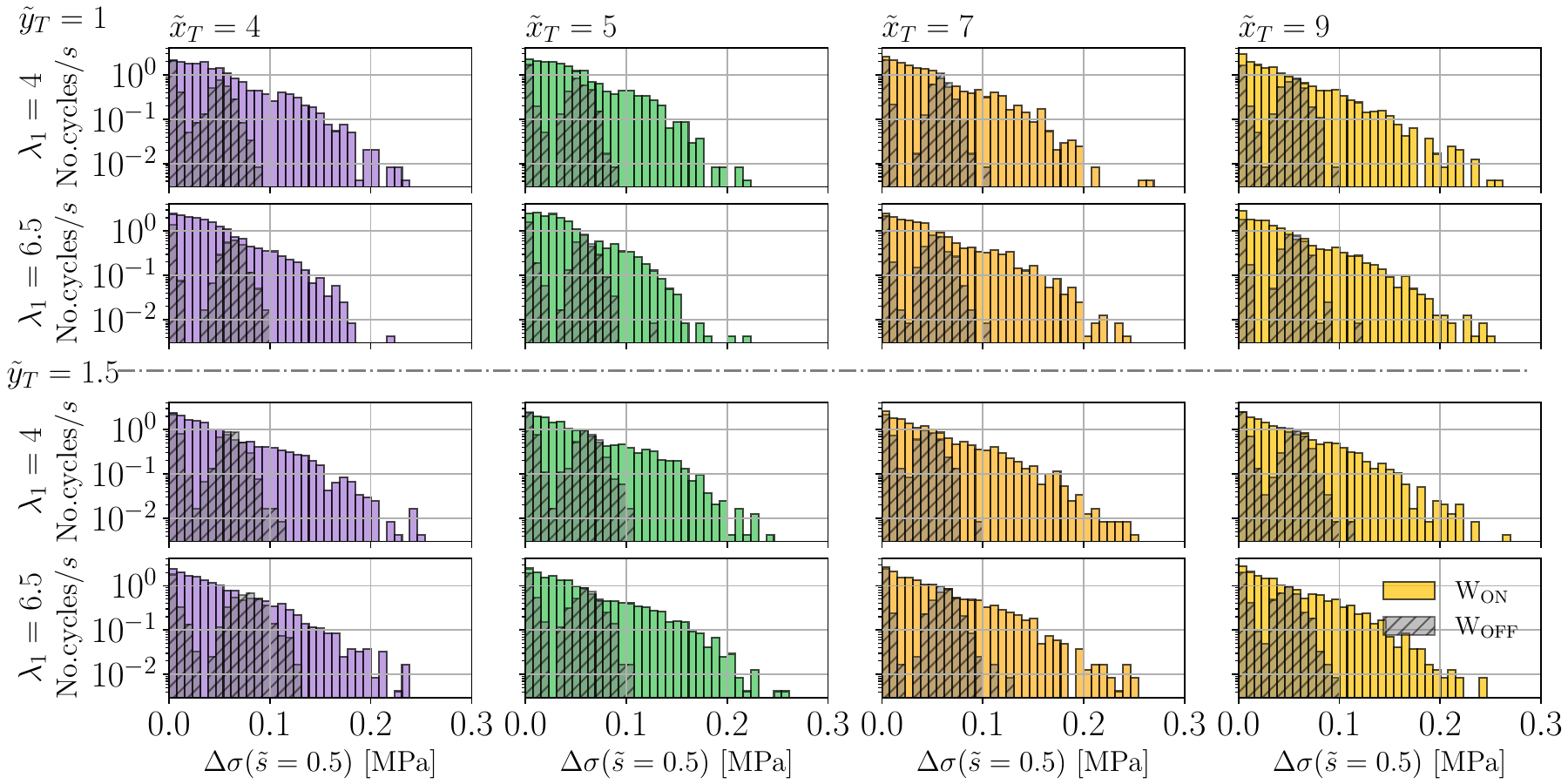}
	\caption{Rainflow cycle amplitude distributions of the flapwise stress $\Delta\sigma(\tilde{s} = 0.5)$ for \enquote{wind-on} ($\mathrm{W_{ON}}$, coloured bars) and \enquote{wind-off} ($\mathrm{W_{OFF}}$, hatched bars) conditions, shown for \textit{a)}: $\tilde{y}_T \in \{0, 0.5\}$ (top and bottom halves), $\tilde{x}_T \in \{2, 3, 5, 9\}$ (columns), and both tip-speed ratios $\lambda_1 \in \{4, 6.5\}$ (rows within each half), and for \textit{b)}:  $\tilde{y}_T \in \{1, 1.5\}$ and $\tilde{x}_T \in \{4, 5, 7, 9\}$.}
	\label{fig:wt3_fig7}
\end{figure}

We focus on the $\mathrm{MIDSPAN}$, specifically at $\tilde{s}=0.5$, where peak mean strain is maximised across the blade's span, to perform a cycle-based analysis of the experienced loading dynamics under different operating conditions. The sensitivity of this choice to the fatigue-loading analysis is discussed below and in the \emph{Supplementary material}.
The bending stress amplitude is estimated from Hooke's law $\sigma = E \varepsilon$,
where $E$ corresponds to the blade's Young's modulus of PLA (assumed to be $E=3~\mathrm{GPa}$---see \citep{Ultimaker}).
The resulting converted stress time series obtained for each of the cases was processed using rainflow counting \citep{amzallag1994} to extract individual stress cycles presented in figure~\ref{fig:wt3_fig7}. The number of cycles in the stress cycle distributions is normalised by the total duration of each test, allowing for a consistent comparison of cycle counts across \enquote{wind-on} and \enquote{wind-off} conditions, being presented in $\mathrm{No.~cycles/s}$. This normalisation is necessary due to the shorter duration of the \enquote{wind-off} measurements and ensures that the reported distributions reflect relative cycle occurrence rates. A clear distinction is observed between \enquote{wind-on} and \enquote{wind-off} cases, with the latter exhibiting significantly fewer cycles across all amplitudes, confirming the dominant role of aerodynamic forcing in driving the experienced blade dynamics.

Under fully waked conditions ($\tilde{y}_T = 0$), the distributions are sharply peaked and concentrated at low stress amplitudes, consistent with the reduced inflow velocity and consequently lower aerodynamic loading. This behaviour is observed for both $\lambda_1$ tested. As the streamwise separation $\tilde{x}_T$ increases, the distributions progressively flatten, reflecting the recovery of the wake and the associated increase in loading magnitude. In this regime, the distributions for $\lambda_1 = 6.5$ remain slightly steeper than those for $\lambda_1 = 4$, in line with the previously observed differences in the streamwise velocity half-wake profiles.

For partial wake conditions ($\tilde{y}_T = 0.5$), we observe a pronounced transition in the first two streamwise stations tested ($\tilde{x}_T \in \{2,3\}$), where the distributions become significantly flatter compared to the fully waked case. This indicates a substantial increase in the occurrence of higher-amplitude cycles, consistent with the intermittent exposure of the rotor to both waked and unwaked flow regions. As $\tilde{x}_T$ increases, the distributions gradually become steeper, suggesting a reduction in intermittency as the wake recovers and the inflow becomes more uniform.

For $\tilde{y}_T = 1$ and $1.5$, the distributions exhibit comparatively minor differences, particularly for $\tilde{y}_T = 1.5$, where the wake influence is minimal and the turbine operates under near free-stream conditions. The distributions in these cases are characterised by elevated baseline amplitudes but a more consistent decay, indicating a more stable loading environment. Some differences between the two tip-speed ratios are observed at intermediate streamwise positions (e.g. $\tilde{x}_T \in \{4,5\}$), where $\lambda_1 = 4$ tends to produce slightly flatter distributions than $\lambda_1 = 6.5$, consistent with the dynamics observed so far.

A relative cumulative fatigue loading index, $D_{rel}$, is calculated by applying rainflow counting and power-law amplitude weighting to the measured stress cycles:
\begin{equation}
D_{rel} = \sum_i n_i (\Delta \sigma_i)^m ,
\label{eq:DREL}
\end{equation}
where $n_i$ denotes the number of cycles associated with a stress range $\Delta \sigma$, and the W\"{o}hler exponent $m$ weights the contribution of high-amplitude events. The term \enquote{cumulative} refers to summation over all identified cycles during the measurement period.
The choice of $m$ sets the relative importance assigned to high-amplitude cycles and may therefore influence the resulting $D_{rel}$ across the blade span. The sensitivity of $D_{rel}$ to $m$ and spanwise measurement location is assessed in the \emph{Supplementary material}.
Note that because the adopted exponent is not a calibrated material S-N parameter, $D_{rel}$ is not interpreted as Palmgren-Miner damage or absolute fatigue-life consumption.
Reported Wöhler exponents for additively manufactured PLA under fully-reversed axial loading are of order $5$--$6.5$ \citep{Ezeh2018PLAFatigue}, itself lower than the higher exponents typical of composite ($m\sim 8$--$12$) wind turbine blade materials \citep{Moens2022a}.
A deliberately low exponent, $m=2$, is used for the primary analysis to test whether the identified partial-wake trend persists without strongly amplifying the contribution of the largest rainflow cycles. The analysis is performed at $\tilde{s}=0.5$, providing a consistent measure of the loading dynamics across the different operating conditions.
By normalising with respect to the case that approximates the best non-waked inflow ($\mathcal{D} = D_{\mathrm{rel}(\tilde{x}_T,\tilde{y}_T)}/D_{\mathrm{rel}(\tilde{x}_T=4,\tilde{y}_T=1.5)}$), $\mathcal{D} > 1$ indicates an increase in relevant loading dynamics due to wake interaction, while $\mathcal{D} < 1$ corresponds to a reduction. We emphasize that $\mathcal{D}$ should therefore be read throughout as a normalised indicator of the wake-induced redistribution of cycle amplitudes relative to the baseline operating point, and not as an absolute fatigue estimate for full-scale machines.

\begin{figure}
	\centering
	\includegraphics[width=.8\textwidth]{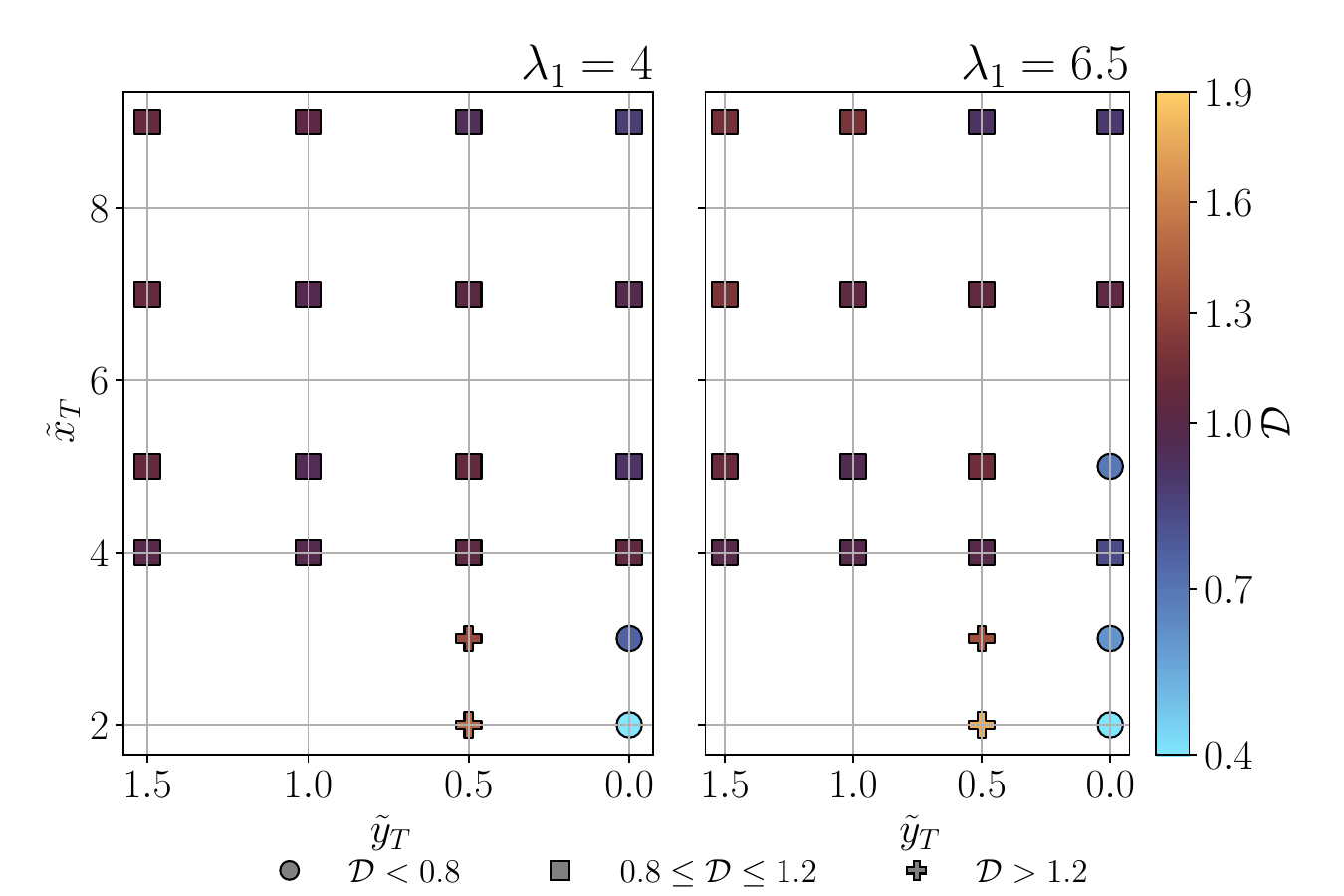}
	\caption{Relative fatigue loading index $\mathcal{D}$ across the $(\tilde{x}_T, \tilde{y}_T)$ experimental matrix for $\lambda_1 = 4$ (left) and $\lambda_1 = 6.5$ (right). Circle markers denote $\mathcal{D} <0.8$, cross markers $0.8\leq\mathcal{D}\leq1.2$, and square markers $\mathcal{D} > 1.2$.}
	\label{fig:wt3_fig8}
\end{figure}

The resulting $\mathcal{D}$ space across $(\tilde{x}_T, \tilde{y}_T)$ is shown in figure~\ref{fig:wt3_fig8}. Different markers are added to help visually identify regions where relative damage is 20\% larger than the reference case, 20\% smaller and within 40\% of the reference case. A clear dependence on $\tilde{y}_T$, especially for $\tilde{x}_T\in\{2,3\}$ is observed. Under fully waked conditions ($\tilde{y}_T = 0$), $\mathcal{D}$ is consistently below unity at small streamwise separations, reflecting the reduced inflow velocity and the associated suppression of high-amplitude stress cycles. The smallest value of $\mathcal{D}$ observed is for $(\tilde{x}_T,\tilde{y}_T)=(2,0)$, in line with the half-wake velocity profiles documented in figure \ref{fig:wt3_fig2}. As $\tilde{x}_T$ increases, $\mathcal{D}$ approaches unity, as the loading progressively returns towards the non-waked condition.

In contrast, half-waked inflow conditions ($\tilde{y}_T = 0.5$) exhibit the largest magnitude of $\mathcal{D}$ for all cases, particularly in the near wake ($\tilde{x}_T \in\{ 2,3\}$), where $\mathcal{D}$ significantly exceeds unity. This is directly related to observations made by \citep{Moens2022a}, where the authors link the direct impact of wake impingement to increased fatigue dynamics of downstream turbines with half-waked inflow conditions producing the most damage. This behaviour is also directly linked to the broadened rainflow distributions observed in figure~\ref{fig:wt3_fig7}, and reflects the enhanced occurrence of large-amplitude cycles generated as the blades periodically sample the highly intermittent boundary of the wake generated by $WT_1$, producing a dominant once-per-revolution (1P) loading cycle. Additional contributions arise from variations in the blade's aerodynamic state—and hence its relative velocity—as it sweeps across these regions, further broadening the rainflow distribution.
For $\tilde{y}_T \geq 1$, $\mathcal{D}$ remains close to unity across all streamwise positions, suggesting that load dynamics approach that of clean inflow conditions. Minor variations are observed between the two tip-speed ratios, but no systematic trend emerges, suggesting that the dominant contribution to load dynamics is governed by the wake interaction rather than the specific operating condition of the upstream turbine. The main difference between the two $\lambda_1$ tested is observed, once more, under half-waked conditions at $\tilde{x}_T=2$. Under these circumstances, for $\lambda_1=6.5$, the turbine's rotor is exposed to an increased variability of impinging mean velocities across the rotor's span than for $\lambda_1=4$ (see figure~\ref{fig:wt3_fig2} \textit{b)}). This potentially contributes to a larger accumulation of damage, resulting in an increased $\mathcal{D}$ at this location ($\tilde{x}_T,\tilde{y}_T = (2,0.5)$).

\begin{figure}
	\centering
	\includegraphics[width=.8\textwidth]{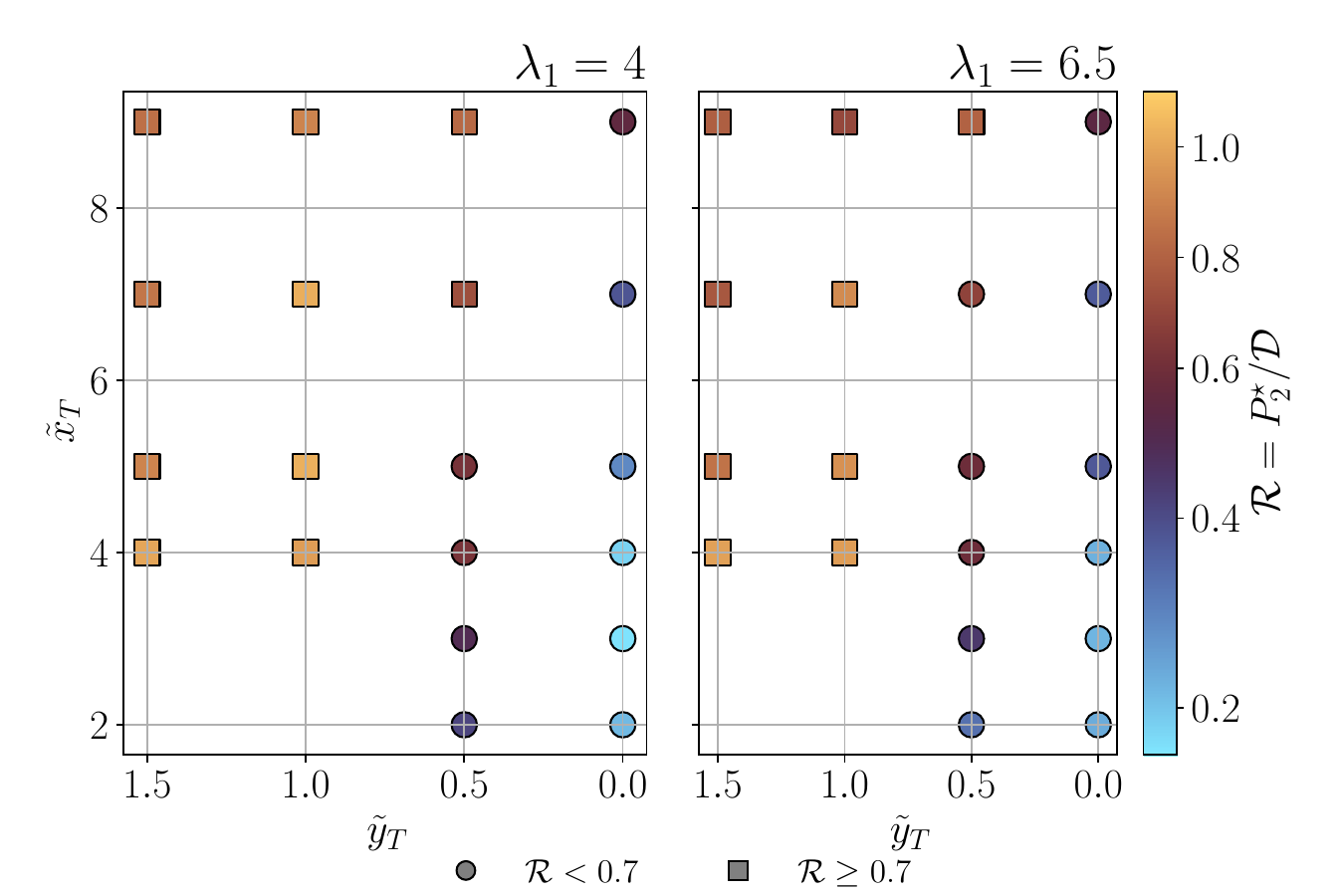}
	\caption{Relative performance-to-loading metric $\mathcal{R} = P_2^{\star}/\mathcal{D}$ across the $(\tilde{x}_T, \tilde{y}_T)$ experimental matrix for $\lambda_1 = 4$ (left) and $\lambda_1 = 6.5$ (right), indicating the trade-off between power production and experienced loading dynamics.}
	\label{fig:wt3_fig9}
\end{figure}

To quantify the trade-off between aerodynamic performance and fatigue-relevant structural loading, we define a normalised aero-structural performance index $\mathcal{R}$, defined as:
\begin{equation}
\mathcal{R}=\frac{P_2^{\star}}{\mathcal{D}^w},
\end{equation}
where $P_2^{\star}$ is the is the downstream-turbine power normalised by the reference condition, $\mathcal{D}$ is the corresponding relative cumulative fatigue-loading index, and $w$ controls the sensitivity of the index to structural loading relative to the unit merit of power production. Values $\mathcal{R}>1$ indicate a more favourable power-to-loading balance than the reference condition, whereas $\mathcal{R}<1$ indicates a less favourable balance. 
The case $w=1$ defines the nominal power-to-loading ratio. Values $w>1$ impose a stronger penalty on fatigue-relevant loading, while $0<w<1$ reduce that penalty. Note that in real life, $w$ can be more complex bearing different cost and maintenance factors not considered here, and this serves as a foundational exercise to weight power production, and relevant load dynamics solely.
By construction, $\mathcal{R} = 1$ corresponds to the baseline non-waked condition. We start by presenting a map of $\mathcal{R}$ for $w=1$ as a function of $\{\tilde{x}_T,\tilde{y}_T\}$ in figure~\ref{fig:wt3_fig9}. Markers are introduced to distinguish operating conditions for which $\mathcal{R}$ remains above 70\% of the reference value. 

The distribution of $\mathcal{R}$ across the $(\tilde{x}_T, \tilde{y}_T)$ space reveals two distinct regimes governed by the lateral wake position. For $\tilde{y}_T \geq 1$, corresponding to near free-stream conditions, $\mathcal{R}$ remains close to or above the $70\%$ threshold across most streamwise positions, indicating a favourable balance between power production and relevant experienced load dynamics. The threshold of $70\%$ was selected following a sensitivity analysis over a range of values, which showed that the qualitative identification of favourable operating regions was largely insensitive to the precise choice of the separating threshold.
In contrast, for $\tilde{y}_T \leq 0.5$, $\mathcal{R}$ is generally reduced, with values in general falling below the $70\%$ threshold. Under fully waked conditions ($\tilde{y}_T = 0$), the reduction in power dominates the respective reduction in experienced loading dynamics, resulting in consistently low $\mathcal{R}$ at small $\tilde{x}_T$, with a gradual recovery as the wake develops downstream. For partial wake conditions ($\tilde{y}_T = 0.5$), $\mathcal{R}$ remains low in the near wake, reflecting the strong increase in experienced loading dynamics due to intermittent loading induced by periodic exposure to waked and non-waked conditions. As $\tilde{x}_T$ increases and the wake recovers, $\mathcal{R}$ exceeds the $\mathcal{R} = 70\%$ threshold, suggesting a general improvement on the operating conditions at which the wind turbine is working.

\begin{figure}
	\centering
	\includegraphics[width=0.8\textwidth]{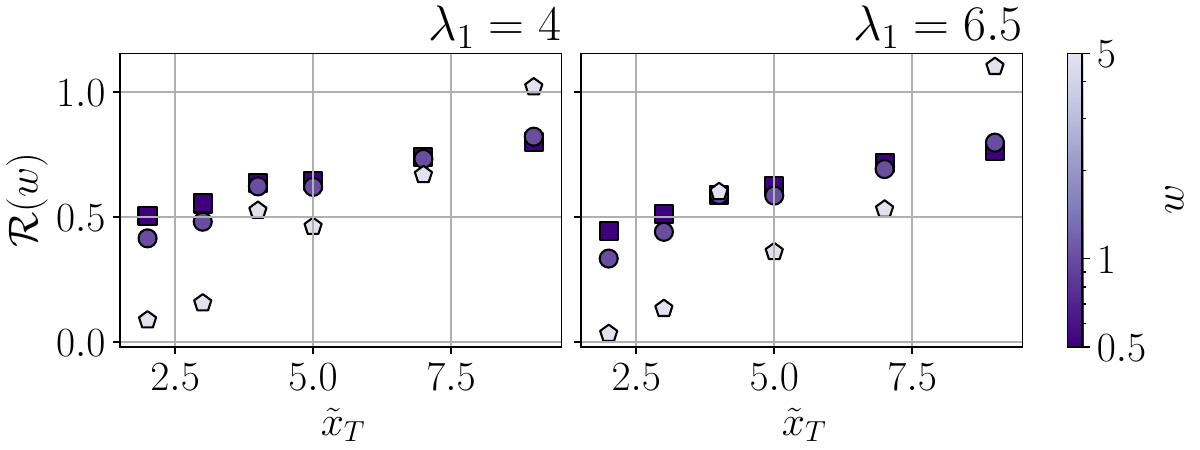}
		\caption{Distribution of $\mathcal{R}(w)=P_2^\star/(\mathcal{D}^w)(\tilde{y}_T=0.5, \tilde{x}_T)$ as a function of $w$.}
	\label{fig:wt3_fig10}
\end{figure}

Figure~\ref{fig:wt3_fig10} illustrates the influence of the weighting parameter $w$ on the combined performance metric for the half-waked condition ($\tilde{y}_T=0.5$). Increasing $w$ progressively assigns greater importance to fatigue loading relative to power production, thereby increasingly penalising operating conditions with larger values of $\mathcal{D}$. Consequently, the near-wake locations ($\tilde{x}_T\in\{2,3\}$) exhibit the strongest reduction in $\mathcal{R}$, whereas the downstream positions remain comparatively insensitive to the choice of $w$. As a result, the separation between favourable and unfavourable operating locations becomes more pronounced as fatigue is prioritised. Nevertheless, for both upstream operating conditions, the downstream spacings consistently provide the highest values of $\mathcal{R}$, demonstrating that the principal conclusions regarding the preferred operating regions are robust to the adopted weighting.

\section{Conclusions}

This paper investigated wake-induced blade dynamics in a two-turbine configuration, focusing on the effect of inter-turbine spacing. The results show that time-averaged loading is primarily governed by the inflow velocity deficit, while unsteady loading depends on both inflow recovery and aerodynamic sensitivity. Fully waked conditions reduce loading due to lower velocities, whereas partially and non-waked conditions recover the aerodynamic response of the blade.

The principal finding of this work is that wake-induced fatigue relevant loading is governed by the cyclic transition between distinct aerodynamic operating states experienced under partial wake overlap. The analysis of the acquired strain dynamics reveals that full and partial wake exposure are not equally detrimental, and should not be treated as such in wind-farm design. Full wake immersion suppresses both mean and fluctuating loads, and its main cost to the downstream turbine is a loss of power production rather than a disproportionate increase in fatigue loading. 
One could naively link the increased velocity fluctuations in the wake of a wind turbine to an increase in respective fluctuating loads, with a respective increase in fatigue accumulation. However, this was not observed.
Partial wake exposure introduces strong intermittency as the rotor transitions between waked and unwaked flow once per revolution, producing the largest occurence of relevant experienced load dynamics to the accumulation of fatigue relevant loading, despite only a partial recovery of power. Near free-stream conditions produce the largest magnitude of mean loads, but comparatively to partial wake inflow conditions, a more stable operating condition limiting extreme fatigue-driving events. 
Free-stream turbulence is likely to actively contribute to the experienced loads and to the accumulation of fatigue damage as observed by \citet{Thomsen1999Fatigue} however, we observe that the change in operating state of the wind turbine rotor induced by a non-homogeneous spanwise mean-velocity profile introduces the largest magnitude of experienced structural dynamics. 
This suggests that the load fluctuations experienced by the blade are influenced less by elevated free-stream turbulence than by the periodic variation in effective tip-speed ratio (hence, the blade's aerodynamic state) as the blade traverses an inhomogeneous velocity field at a fixed rotational speed.

By combining power output and the experienced load dynamics into the metric $\mathcal{R} = P_2^\star/\mathcal{D}^{w}$, we show that partial wake conditions yielded the lowest aero-structural performance index, indicating the least favourable balance between power production and fatigue-relevant cyclic loading. This distinction highlights that minimising partial wake overlap potentially driven by wind-turbine placement or, by wake-steering techniques should be an explicit consideration in wind-farm layout strategies. This conclusion is robust to the choice of $w$, as the sensitivity analysis shows that the relative performance of the different operating conditions remains qualitatively unchanged as greater weight is assigned to fatigue-relevant loading.

These conclusions are drawn from small-scale, model-turbine measurements ($Re_D\approx2\times10^5$) and should be interpreted with the corresponding caveats. The blades were manufactured from PLA, whose fatigue behaviour differs from the composite laminates used in full-scale rotors, and the relative damage metric $\mathcal{D}$ is intended as a trend indicator rather than an absolute fatigue estimate (see discussion following equation~\ref{eq:DREL}). The hot-wire characterisation of the inflow was limited to the streamwise velocity component, so the role of wake swirl in the observed loading asymmetry remains unresolved. Extending this framework to higher Reynolds numbers, composite blades and fully three-component inflow measurements would establish how these experimentally identified fatigue mechanisms translate to utility-scale wind farms and future wake-aware control strategies.

\begin{Backmatter}

\paragraph{Acknowledgements}
We gratefully acknowledge the help of Paul Howard, Kevin Gouder, William McArdle, Ricardo Huerta and Dominica Rhozinski.

\paragraph{Funding Statement}
F. J. G. de Oliveira and O. R. H. Buxton wish to acknowledge financial support given by EPSRC through grant no. EP/V006436/1. Use was also made of EPSRC grant no. EP/L024888/1 for access to the National Wind Tunnel Facility (NWTF).

\paragraph{Declaration of Interests}
The authors declare no conflict of interest.

\paragraph{Data Availability Statement}
Raw data are available from the corresponding author (F. J. G. de Oliveira).

\paragraph{Ethical Standards}
The research meets all ethical guidelines, including adherence to the legal requirements of the study country.

\end{Backmatter}

\bibliographystyle{unsrtnat}
\bibliography{bibliography}

\end{document}